\documentclass{edp-conf}
\usepackage{graphicx}
\begin{document}

\TitreGlobal{SF2A 2003}

\title{Stellar populations and their kinematics from high and medium
resolution spectra: mixed inversions}
\author{P. Ocvirk $^{1}$, A. Lancon $^{1}$, C. Pichon}\address{Observatoire de Strasbourg,
$^{2}$ CRAL, $^{3}$ Institut d'Astrophysique de Paris, $^{4}$  Observatoire de Bordeaux}
\author{P. Prugniel $^{2}$, E. Thiebaut $^{2}$}
\author{D. Le Borgne $^{3}$, B. Rocca-Volmerange $^{3}$, M. Fioc $^{3}$}
\author{C. Soubiran $^{4}$}
\runningtitle{Inversions of galaxy spectra}
\setcounter{page}{237}
\index{Ocvirk, P.}
\index{Lancon, A.}
\index{Lancon, A.}
\index{Pichon, C.}
\index{Prugniel, P.}
\index{Thi\'{e}baut, E.}
\index{Thiebaut, E.}
\index{Le Borgne, D.}
\index{Rocca-Volmerange B.}
\index{Fioc, M.}
\index{Soubiran, E.}

\maketitle
\begin{abstract} 
We present inversion techniques aiming at recovering the stellar content
(age, metallicity, extinction) and kinematics of a stellar population from its
absorption line spectrum. These techniques use new synthetic high
resolution spectra produced by PEGASE\_HR 
and minimization algorithms. Application to SDSS data and to a simulated disc-bulge system are shown.
\end{abstract}
%
\section{Introduction}
The high resolution spectrographs now installed on 10m-class telescopes open new perspectives in the exploration of the formation of galaxies. This poster presents the first attempt to fit simultaneously the local line-of-sight velocity distribution (LOSVD) and the nature of the stellar population. It is based on efficient non-parametric inversion procedures.
\section{Models of galaxy spectra}
PEGASE\_HR (Fioc \& Rocca-Volmerange 1997; Le
Borgne et al., 2002) produces high resolution synthetic spectra of evolved Single Stellar
Populations (SSPs) assuming a given IMF and stellar evolution tracks. We build
a basis of such spectra by computing SSPs for a set of ages and metallicities. A model galaxy spectrum is a linear combination of the components of
this basis obscured by a dust extinction model and convolved by a
velocity distribution. The model galaxy spectrum thus depends on the following
4 sets of
unknowns: the weight of each component, the metallicity of each component,
the color excess seen by the component and finally the global LOSVD of the
whole population.
\section{Inversion of galaxy spectra}
Inverting an observed spectrum involves recovering the 4 sets of
unknowns. We follow a maximum a posteriori likelihood formulation, assuming
gaussian errors. This inversion problem thus turns into minimizing a
chi-square with a quadratic penalization. Simulations have been performed to investigate the behaviour of the method for several
pseudo-observational contexts.
\section{Application to SDSS data}
\begin{figure}[h]  
\centerline{
\resizebox{2.5cm}{2.5cm}{{\includegraphics{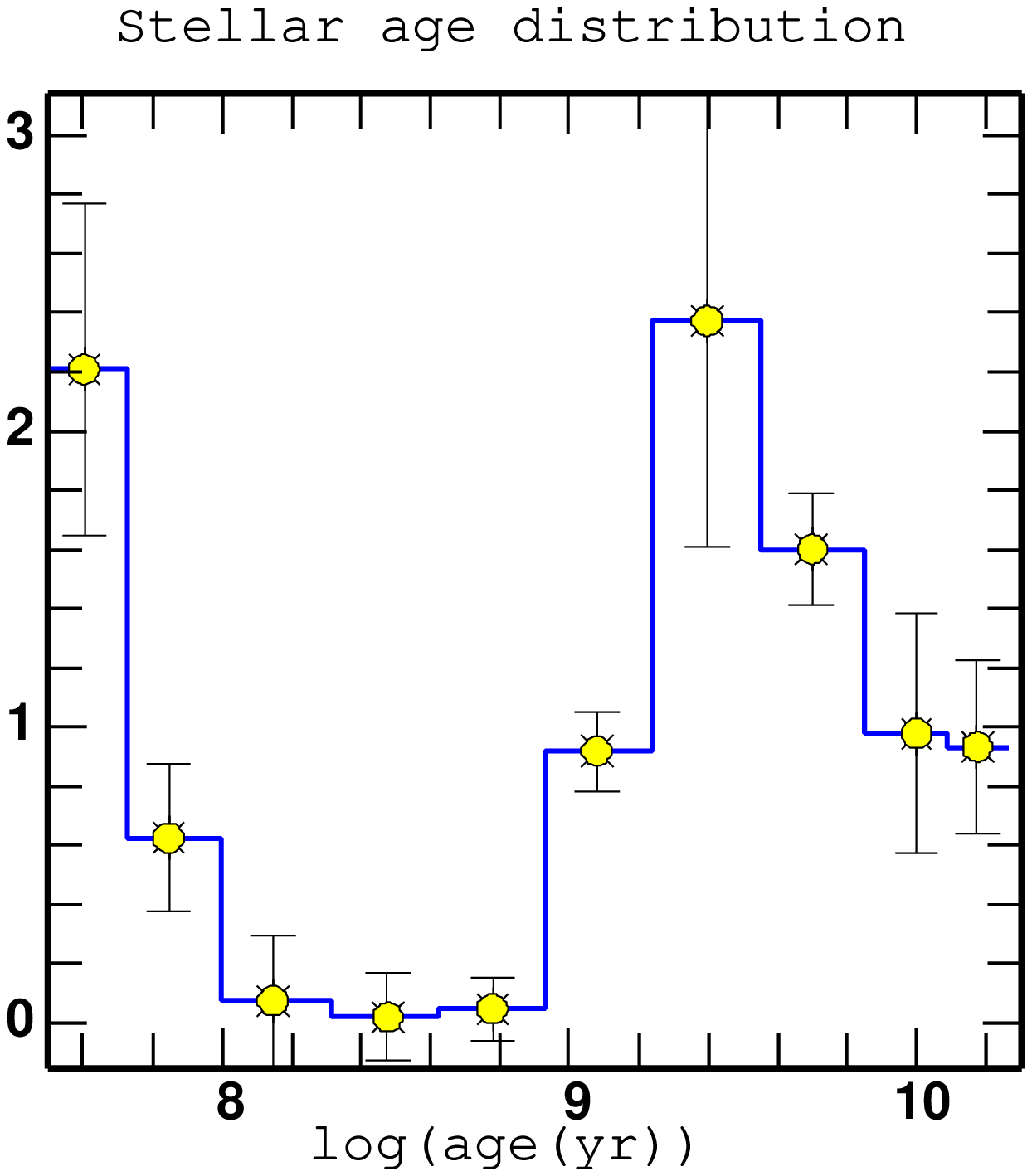}}}
\resizebox{2.5cm}{2.5cm}{{\includegraphics{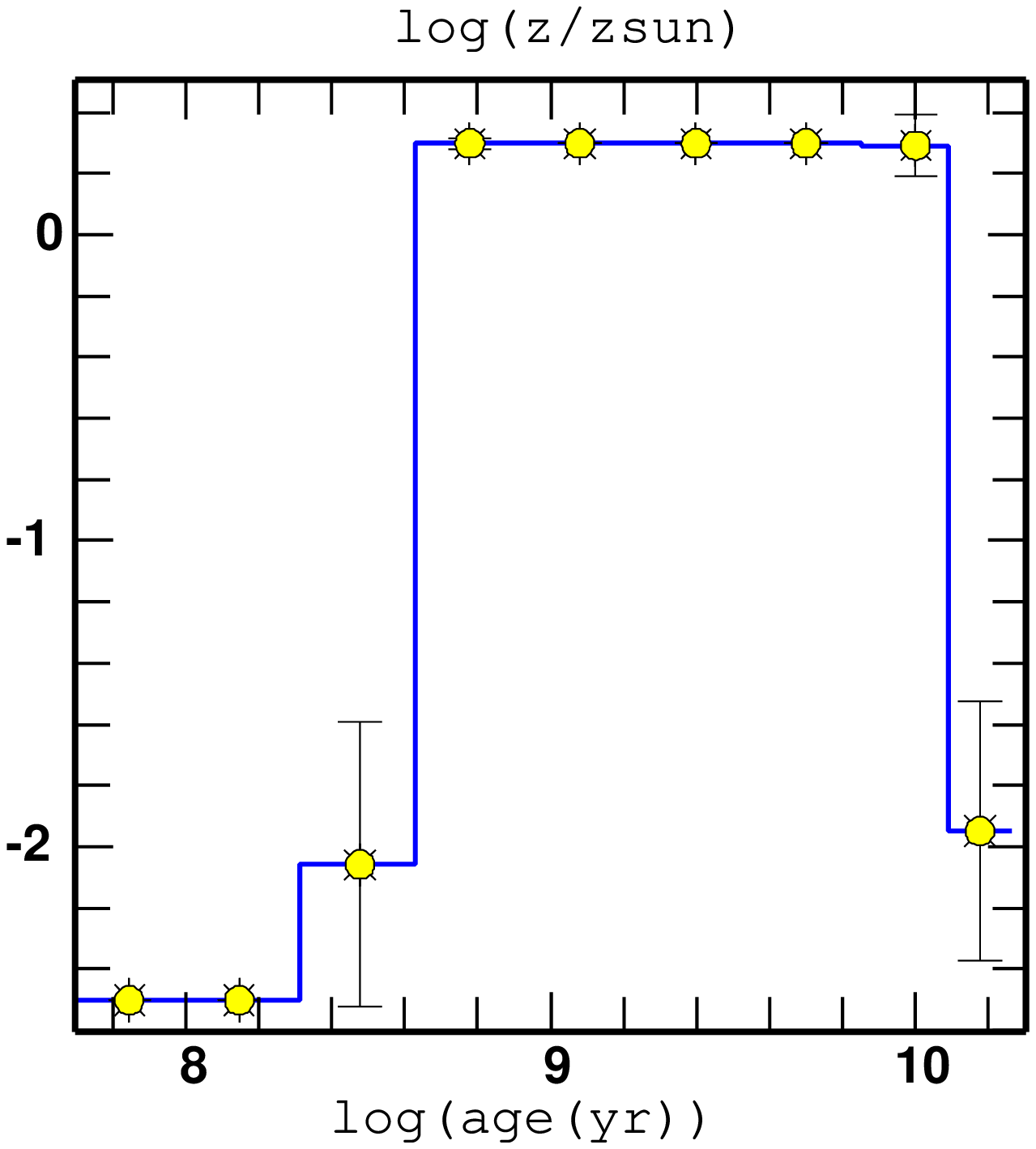}}}
\resizebox{2.5cm}{2.5cm}{{\includegraphics{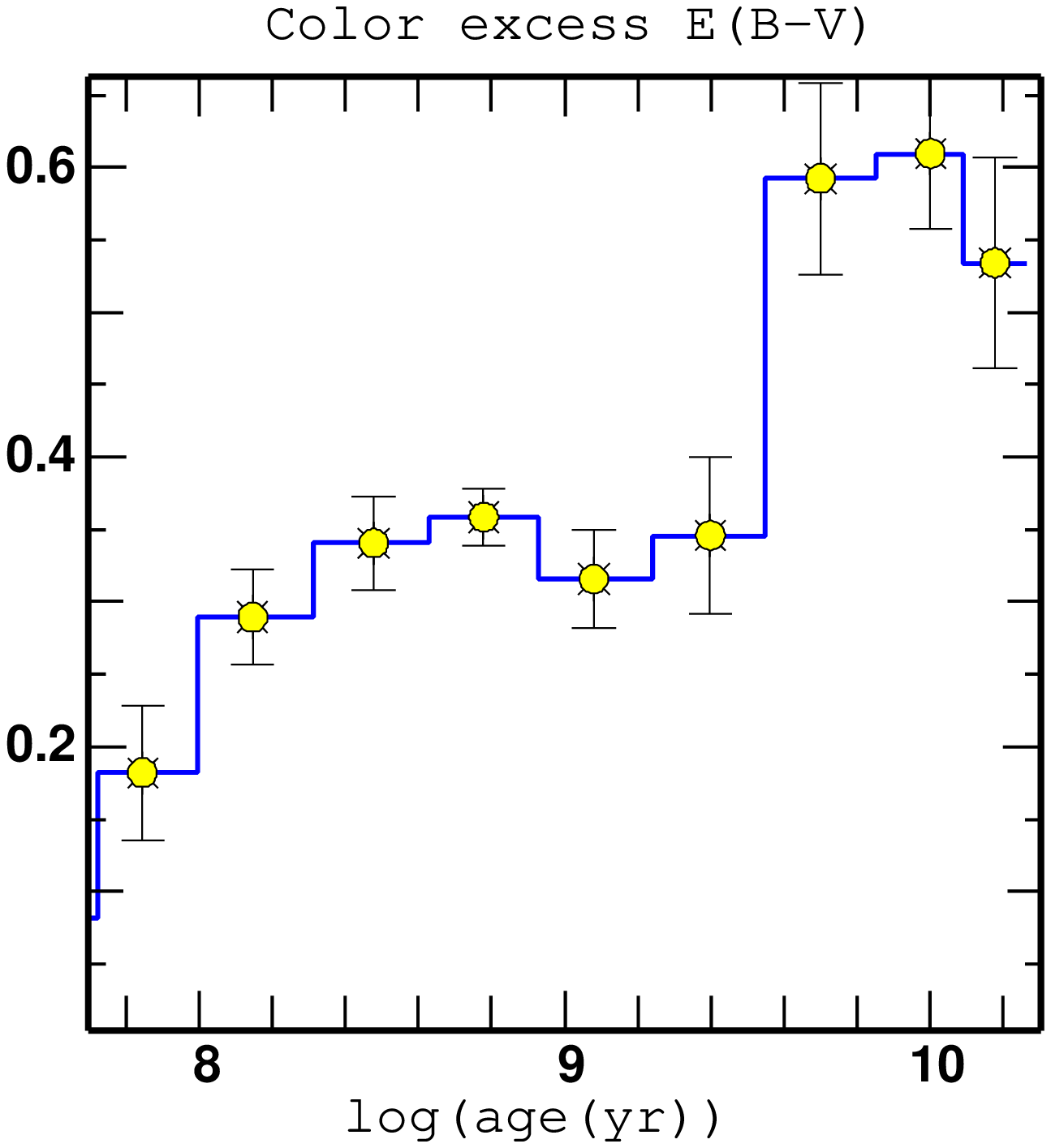}}}
\resizebox{2.5cm}{2.5cm}{{\includegraphics{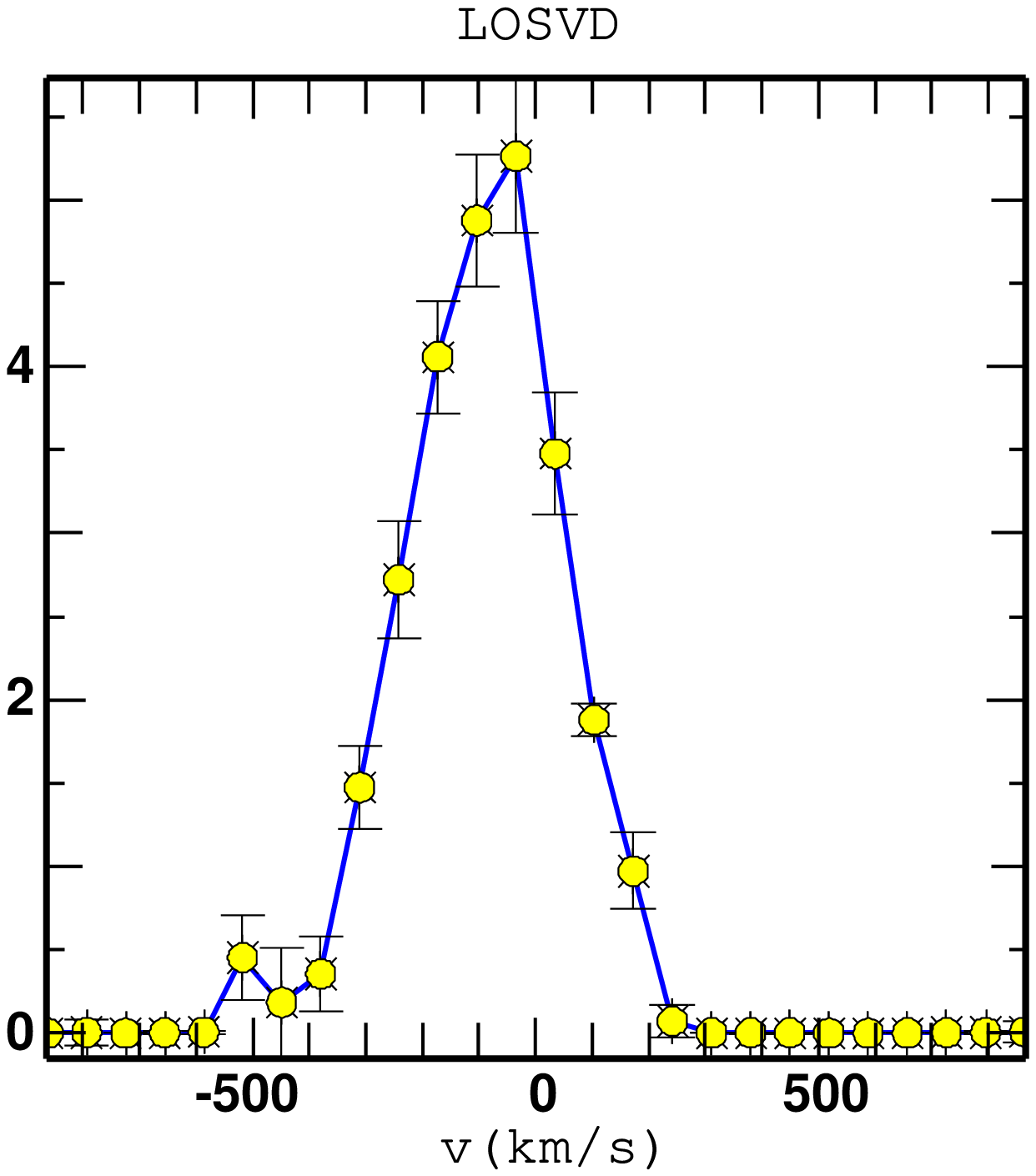}}}
\resizebox{2.5cm}{2.5cm}{{\includegraphics{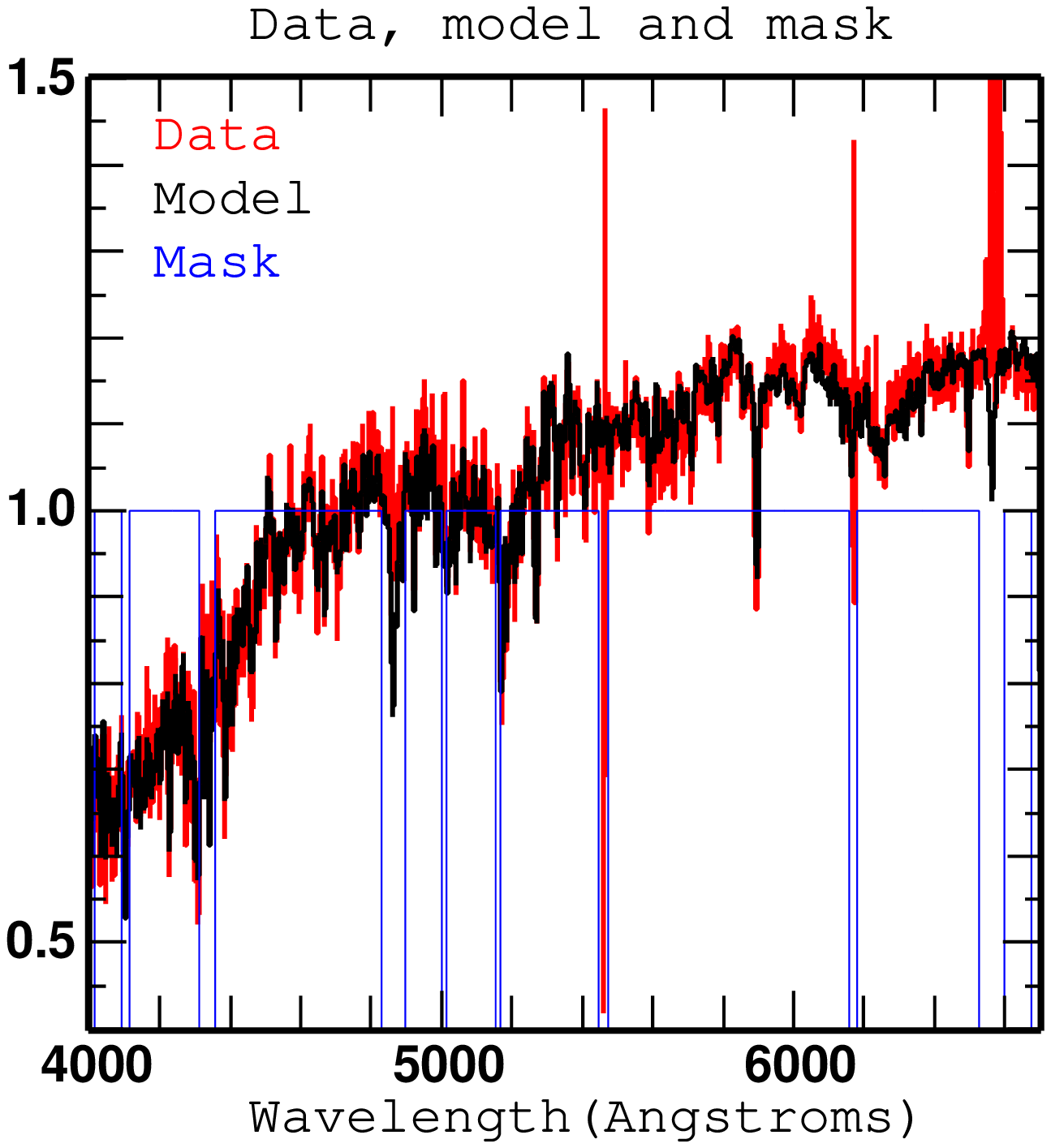}}}}
\caption{ Results of the inversion of the spectrum of the galaxy
SDSS-51630-0266-001. }  
\label{f:sdss}
\end{figure}

As an example, the stellar
content, reddening and kinematic information have been reconstructed for the galaxy
SDSS-51630-0266-001 and shown in figure \ref{f:sdss}. Application to a larger
systematic subset of SDSS data is on course.
\section{Application to disc-bulge separation}
\begin{figure}[h]  
\centerline{
\resizebox{3.3cm}{3.3cm}{{\includegraphics{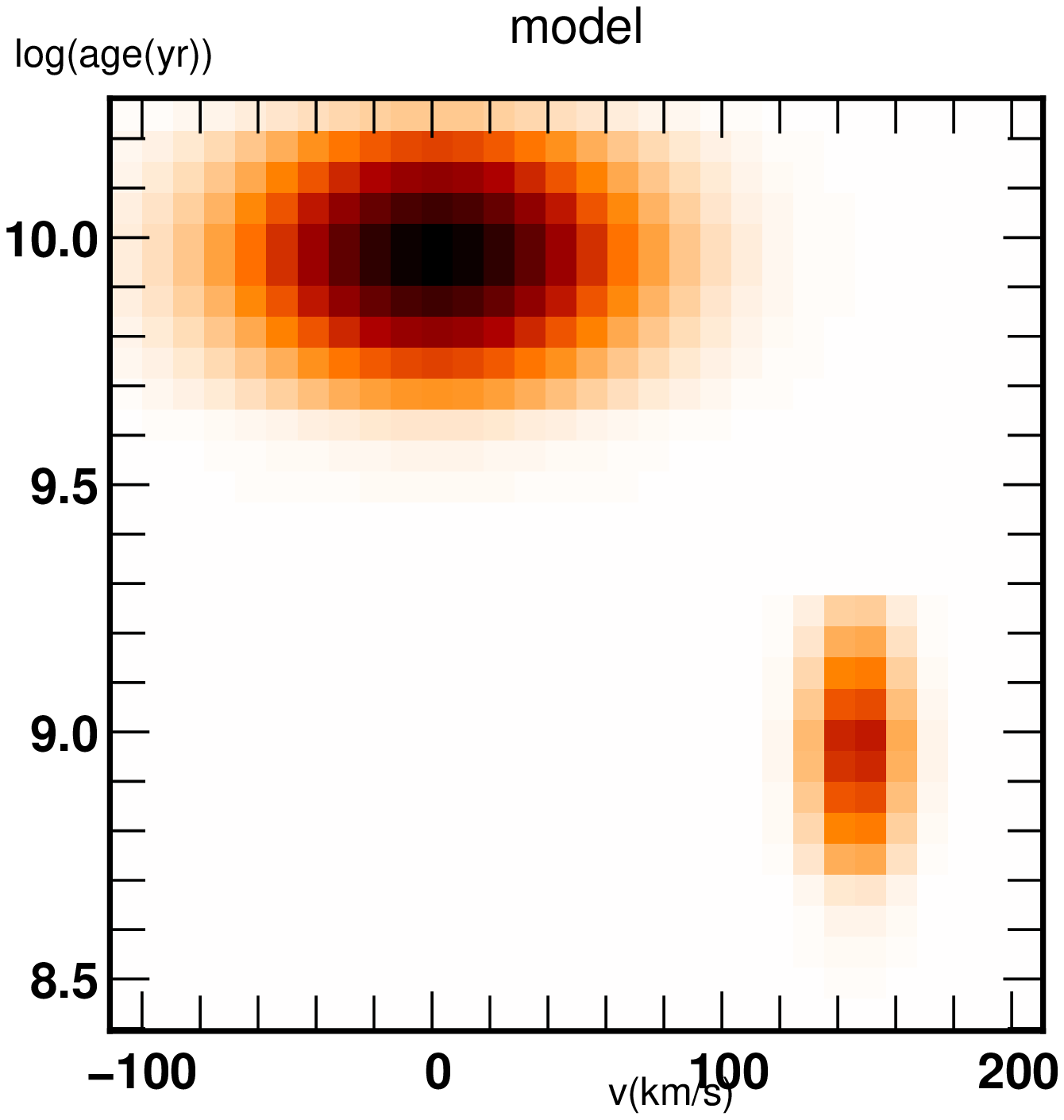}}}
\resizebox{3.3cm}{3.3cm}{{\includegraphics{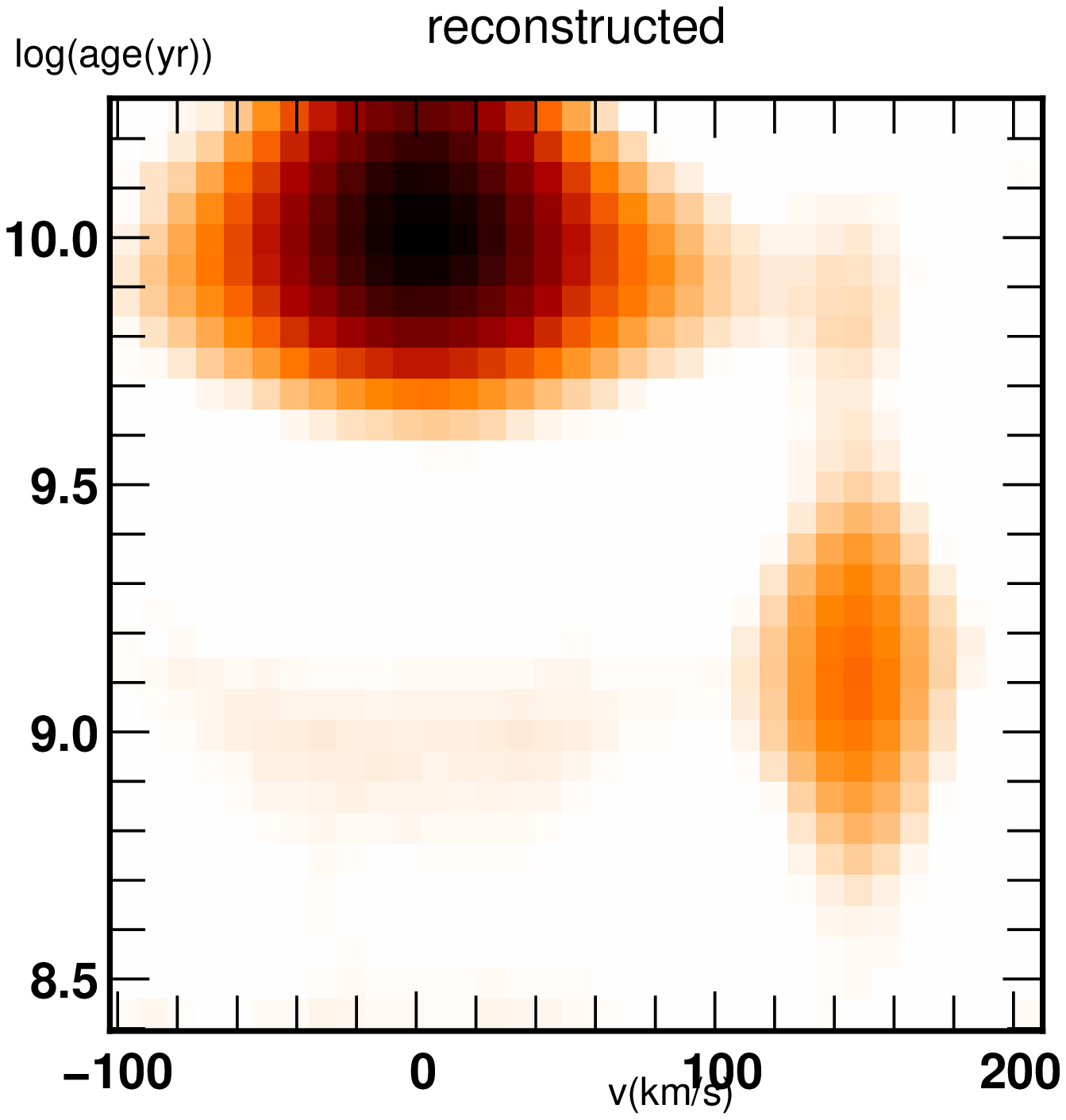}}}}
\caption{ Model and reconstruction of a simulated disc-bulge system. }  
\label{f:cki}
\end{figure}

We have studied the feasibility of separating the disc light from the bulge
light in an idealized case where the disc and bulge stellar populations are
different in SFH and kinematics, mono-metallic and seen without
extinction (figure \ref{f:cki}).


\end{document}